\documentclass{ws-ijmpa}

\begin{document}

\markboth{Tetsuo Hatsuda}
{HADRONS ABOVE $T_c$}

%%%%%%%%%%%%%%%%%%%%% Publisher's Area please ignore %%%%%%%%%%%%%%%
%
\catchline{}{}{}{}{}
%
%%%%%%%%%%%%%%%%%%%%%%%%%%%%%%%%%%%%%%%%%%%%%%%%%%%%%%%%%%%%%%%%%%%%

\title{HADRONS ABOVE $T_c$}

\author{\footnotesize TETSUO HATSUDA }

\address{Department of Physics, University of Tokyo, \\
Tokyo, 113-0033, Japan }

\maketitle

\pub{Received (Day Month Year)}{Revised (Day Month Year)}

\begin{abstract}
Hadronic modes in the quark-gluon plasma and their
spectral properties are discussed on the basis of the 
 lattice QCD data.

\keywords{quark-gluon plasma; lattice QCD; charmonium.}
\end{abstract}

\section{Introduction}

 Fate  of light hadronic excitations above the 
 critical temperature $T_c$ of the QCD phase transition
 was first investigated  20 years ago.\cite{HK85,De85} 
 
 In Ref.~\refcite{HK85} whose title is 
 ``Fluctuation effects in hot quark matter: precursors of 
 chiral transition at finite temperature", the hadrinic modes 
  above $T_c$ were discussed in connection with 
  the critical fluctuation
  associated with chiral symmetry restoration:
 The abstract says ``$\cdots$ {\it there arise soft modes having
 a large strength and a narrow width above
 the critical temperature,
 which are analogous to the fluctuation of the 
 order parameter in a superconductor above the 
 critical point.}"   
  An example of such soft mode ($\sigma$-$\pi$) at $T > T_c$
  was demonstrated 
  using  the Nambu-Jona-Lasinio model\cite{HK85,review}
   and was later called  the ``para-pion".\cite{Ha95}
   From the modern point of view, It could be identified as 
   a pre-formed $q\bar{q}$ resonance  in the pseudo-gap phase
    where the chiral order parameter has  a large fluctuation above $T_c$
  (see e.g. Refs.~\refcite{BB00,KKN}).
  
   In Ref.~\refcite{De85} whose title is 
   ``Conjecture concerning the modes of excitation of the quark-gluon plasma'',
    the hadronic modes above $T_c$ were discussed
   with a different perspective which is unique in non-Abelian gauge theories,
    i.e. the dynamical confinement of the plasma above $T_c$: The 
    abstract says
  `` $\cdots$ {\it the plasma exhibits confining features similar to 
 that of the low-temperature hadronic phase. The 
 confining features are manifest in the long-range,
 i.e., long-wavelength, low-frequency, modes 
 of the plasma.}'' 
 
  Recently, studies of the hadronic modes in the time-like domain
   become possible in lattice QCD simulations owing to the 
   technique of the maximum entropy method (MEM).\cite{NAH99,AHN01}
  It was then found that some hadronic modes 
  (in the $s\bar{s}$ sector) survive even at $T > T_c$
   in quenched  simulations  (see Sec. \ref{sec:anisotropic}).\cite{AHN03} 
   However,
   its relation to the above ideas (chiral pseudo-gap or dynamical
    confinement) is not yet clear.

 For heavy hadrons such as the  charmonia,
   it has been considered that they dissolve 
  soon after the deconfinement because the 
  screened Coulomb potential above $T_c$ is not 
   strong enough to hold the $c\bar{c}$ pair.\cite{HHKM86,MS86}
 However,  the quenched lattice QCD simulations\cite{AH04,Umeda,Datta}
  have recently shown that resonances such as
   $J/\Psi$ and $\eta_c$ survive even up to $T/T_c \sim 1.6$
    (see Sec. \ref{sec:anisotropic}). Physical origin of 
    this phenomena is also an open question at the moment.

\section{QCD spectral functions and MEM}
\label{sec:1}
     
  The Matsubara correlation  in a 
   mixed representation is defined as 
 $    D(\tau,{\bf p}) = \int  d^3x \ {\cal D}(\tau, {\bf x}) 
     {\rm e}^{-i {\bf {\small p}} \cdot {\bf {\small x}} } , $    
   where $ {\cal D}(\tau, {\bf x})$ is 
    the imaginary time two-point function.
   The spectral representation tells 
\begin{eqnarray}
 \label{eq:7.mixed-sum}
 D ( \tau , {\bf p} ) 
 = \int_{-\infty}^{+\infty} {{\rm e}^{-\tau \omega} \over
 1 \mp {\rm e}^{-\beta \omega}} \ A (\omega, {\bf p})
 \ d\omega \ \ \ (0  \le   \tau < \beta),  
\end{eqnarray}
 which  is  convergent for
  $\tau \neq 0$ in QCD. $A(\omega,{\bf p})$ contains all the information of
 hadronic modes (mass, width etc) in the relevant channel.

Reconstructing a continuous function $A(\omega, {\bf p})$ from
 discrete and finite lattice data $D(\tau,{\bf p} )$ 
 is a typical ill-posed problem.   One way to avoid this difficulty is to introduce 
 an ansatz for the spectral function with a few parameters.\cite{HNS93}
  The maximum entropy method (MEM)
   provides an alternative and much more powerful approach\cite{AHN01,JG96}:
   one can obtain a unique $A$ from the lattice data $D$
   without making  a priori parameterizations.
  Also,  statistical significance of the resultant $A$ can be evaluated.
  
 In MEM,  the most probable
$A$ given lattice data $D$ is obtained by
maximizing the conditional probability 
$P[A|D] \propto  e^{\alpha S - L},$
where $L$ is the standard likelihood function and
$S$ is the Shannon-Jaynes information entropy:
\begin{eqnarray}
S = \int_0^{\infty} \left [ A(\omega ) - m(\omega )
   - A(\omega)\log \left ( \frac{A(\omega)}{m(\omega )} \right ) \right ]
   d\omega . 
  \label{SJ-entropy}
\end{eqnarray}
The statistical significance (error)  of the resultant $A$
is estimated by the second variation, $(\delta/\delta A)^2 P[A|D]$.
The default model $m(\omega)$ in Eq.(\ref{SJ-entropy})
may be chosen so that the MEM errors become minimum.    
The final result is given by
a weighted average over the parameter $\alpha$ as
$A (\omega,{\bf p}) = \int A_{\alpha}(\omega,{\bf p})  P[\alpha|D m] d\alpha ,$
where $A_{\alpha}(\omega,{\bf p})$ is obtained by
minimizing $P[A|D] $ for a fixed $\alpha$. 

  First successful application of MEM
  to the lattice QCD data at $T=0$ was carried out  for the ground and 
  excited mesons in Ref.~\refcite{NAH99}.
  Also, basic concepts and techniques of MEM applied to 
  lattice QCD are summarized  in Ref.~\refcite{AHN01}. 
 
 \section{Applications of MEM ($T=0$)}
  
\begin{figure}
\centerline{\psfig{file=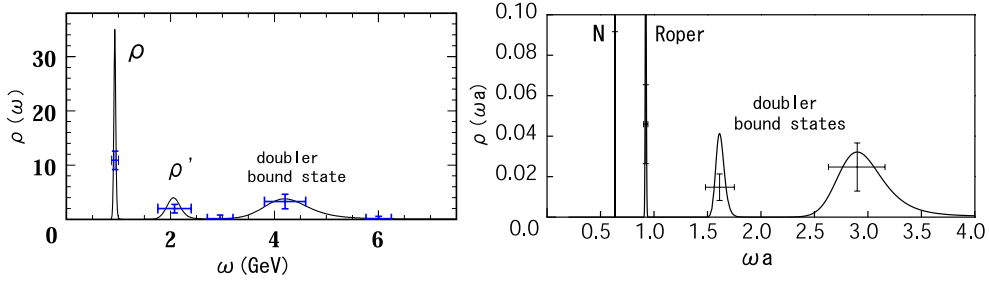,width=13cm}}
\vspace*{8pt}
\caption{Left panel: The dimensionless spectral function 
 $\rho(\omega) = A(\omega, {\bf 0})/(3\omega^2)$
 of the vector meson
 extracted from a $20^3 \times 24$ lattice 
with $\beta =6.0$.
 The hopping parameter is take to be 
  $\kappa=$ 0.1557.  The figure is adapted from Ref.~7. 
  Right panel:
  The dimensionless spectral function $\rho(\omega) = A(\omega,{\bf 0})/\omega^5$
 in the nucleon channel extracted from a $32^4$ lattice 
  with $\beta=6.0$ ($ a \simeq 0.093$ fm).    The hopping parameter is take to be 
  $\kappa=$ 0.1550.
   The horizontal axis denotes a dimensionless frequency
 $\omega a$. This figure is adapted from Ref.~18.}
  \label{fig:RHO-ROPER} 
\end{figure}

  Shown in the left panel of Fig.~\ref{fig:RHO-ROPER} is a  spectral image
   of the vector meson at rest (${\bf p}=0$)
  extracted from the quenched  QCD data.\cite{NAH99}
   The first (second)  peak corresponds to 
   the ground (excited) vector meson. On the other hand,
   the highest peak corresponds to 
   a bound state of Wilson doublers as first pointed out in Ref.~\refcite{Ya02}.
  Shown in the right panel of Fig.~\ref{fig:RHO-ROPER} is the spectral function in the 
   nucleon channel extracted from the quenched  QCD
    data.\cite{SSH05}
    The first (second) peak corresponds to  
  the nucleon and the Roper resonance, while the higher two
   peaks correspond to the bound states of Wilson doublers.
  
\section{Applications of MEM ($T \neq 0$)}
\label{sec:anisotropic}

In Refs.~\refcite{AHN03,AH04}, 
 quenched QCD simulations  with $\beta=7.0$
on $32^3\times N_\tau $  anisotropic lattice were performed
 using naive plaquette gauge action and the standard
Wilson quark action. The renormalized anisotropy is
$\xi = a_{\sigma}/a_{\tau}=4.0$ with 
 $ a_\tau = {a_{\sigma}}/{4} = 9.75\times 10^{-3}\ {\rm fm}$ 
and  $L_{\sigma} = 1.25 $ fm. Employing the
anisotropic lattice is necessary to keep the number of 
 temporal data points as large as possible;
  $N_\tau$ is taken to be 96, 54, 46, 44, 42, 40 and 32 
 which correspond to
  $T/T_c$ = 0.78, 1.38, 1.62, 1.70, 1.78, 1.87 and 2.33, respectively. 

\vspace{0.2cm}

\centerline{\bf $s\bar{s}$ mesons at $T\neq 0$}

\begin{figure}
\centerline{\psfig{file=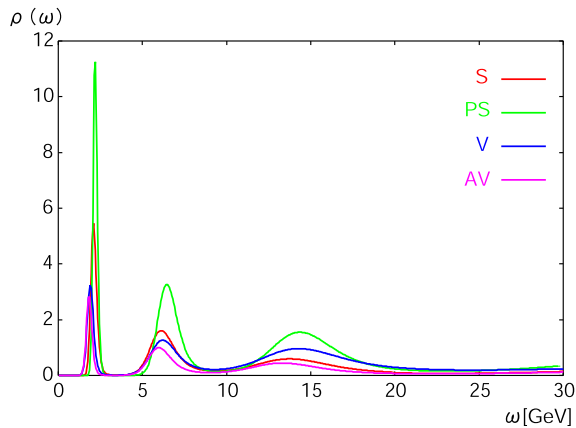,width=7.5cm}}
\vspace*{8pt}
\caption{The dimensionless spectral function 
 measured on an anisotropic lattice for the $s\bar{s}$ mesons at $T/T_c=1.38$.
 The figure is taken from Ref.~9.}
\label{fig:SPF2}    
\end{figure}

 Shown in Fig.\ref{fig:SPF2} is the spectral functions of the 
 $s\bar{s}$ mesons in the scalar (S), pseudo-scalar (PS),
  vector (V) and axial-vector (AV) channels
  for  $T/T_c=1.38$.\cite{AHN03}
  The quark mass is chosen to reproduce the 
  experimental $\phi$ meson mass at $T=0$ approximately.
  One finds that spectral functions in all channels have
 degenerate peaks around 2.5 GeV at $T/T_c=1.38$
  (2.4 times the $\phi$-meson mass
  at $T=0$). Also, these  peaks
 have been shown to 
 disappear at higher temperatures e.g. at $T/T_c=1.87$.\cite{AHN03}
 
\vspace{0.2cm}

\centerline{\bf $c\bar{c}$ mesons at $T \neq 0$ }

\begin{figure}
\centerline{\psfig{file=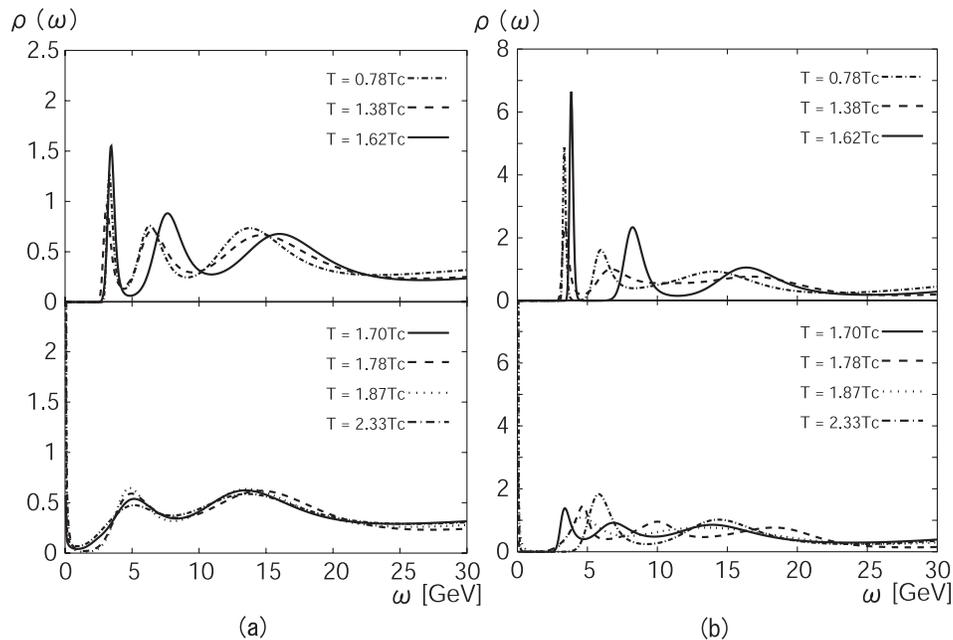,width=12.5cm}}
\vspace*{8pt}
\caption{Spectral functions
 measured on an anisotropic lattice for (a) the $J/\Psi$ channel 
 ($\rho(\omega)=A (\omega,{\bf 0})/(3\omega^2)$)
 and for (b) the $\eta_c$ channel ($\rho(\omega) = A(\omega,{\bf 0})/\omega^2$).
  Figures adapted from Ref.~12.}
\label{fig:SPF}    
\end{figure}

 Shown in  Fig.\ref{fig:SPF} is the 
 case for charmonia.\cite{AH04} 
 Quark masses are chosen so that  we have 
$m_{J/\psi}^{\rm lat}\simeq 3.10 {\rm GeV}$ and $m_{\eta_c}^{\rm lat}\simeq 
3.03 {\rm GeV}$ at $T=0$, 
which should be compared with the experimental values, 3.10 GeV and 2.98 GeV, respectively.

 Sharp peaks for $J/\Psi$ and $\eta_c$ are found 
even up to $T\simeq 1.62T_c$ as shown in Fig.\ref{fig:SPF} with the 
 peak position almost the same as their $T=0$ values.
  Similar observations on the  charmonium bound states
  above $T_c$ were obtained in other independent studies.\cite{Umeda,Datta}
   In our case, 
 the sharp peaks disappear suddenly between  $ 1.6 T_c $ and $1.7 T_c$. 
 The width of the first peak in Fig.\ref{fig:SPF} partly
reflects the unphysical broadening due to
the statistics of the lattice data and
partly reflects possible physical broadening
at finite $T$.  The second and third peaks
in Fig.\ref{fig:SPF} are likely to be related to Wilson doublers.

\vspace{0.2cm}

\centerline{\bf reliability of the MEM images}  

 Lack of
 accurate lattice data can cause
 fake peaks and/or  fake smearing 
  as demonstrated by the mock data in Ref.~\refcite{AHN01}.
  Therefore the spectral functions obtained from MEM
   should pass some reliability tests.
  The first test is 
 the error analysis of the peaks from the 
 second variation, $\delta^2 P[A|D]/\delta A(\omega) \delta A(\omega')$.
 The second test is the change of the spectral functions
 under the variation of the number of data point $N_{\rm data}$ 
 employed for the MEM analysis.
 The third test  is a study of the 
 finite volume effect in the spatial direction.
 The results in Fig.\ref{fig:SPF} have already
  passed the first and second tests. An attempt 
   toward   the third test is recently reported 
 in Ref.~\refcite{iida}.

\vspace{0.2cm}

\centerline{\bf effects of dynamical quarks}

 If dynamical quarks are included,
   there arise two competing effects to the 
    dissociation rate of the hadronic modes observed 
     in the quenched simulations: 
  enhancement due to 
    collisions with dynamical quarks, and
    suppression due to small 
   critical temperature in full QCD
 ($T_c^{\rm quench} \sim 270 \ {\rm MeV}
  \ {\rm vs.}\ T_c^{\rm full} \sim 170 \ {\rm MeV}$).
 The number ratio of the plasma constituents
 between full QCD ($N_f=2$) and quenched QCD ($N_f=0)$ reads 
\begin{eqnarray}
\frac{n_{\rm q+g}(T_c^{\rm full})}{n_{\rm g}(T_c^{\rm quench})}
= \frac{16+21}{16} \cdot \left( \frac{T_c^{\rm full}}{T_c^{\rm quench}} \right)^3
\simeq 0.62.
\end{eqnarray}
 Therefore, 
  the net thermal dissociation rate of the hadronic
  resonances may be  even smaller in full QCD than that in quenched QCD
   for given $T/T_c$. In any case, 
 it is truly necessary to carry out  full QCD simulations to
  make definite conclusions.  A preliminary study in $N_f=2$ QCD
   is recently reported in Ref.~\refcite{alt}.
  
\section{Concluding remarks}

In this report, we have discussed the old idea of hadronic
 modes above $T_c$ (chiral pseudo-gap\cite{HK85}
  and dynamical confinement\cite{De85})
  and  recent quenched lattice QCD results 
  showing the  existence of resonance structures above $T_c$ 
  in $s\bar{s}$ and $c\bar{c}$ channels. 
 To unravel the true nature of such hadronic modes above $T_c$,
  it is important to extract more 
  information from lattice QCD data, such as 
  the spatial wave functons,\cite{QCDTA01}  spectral functions
   with finite spatial momentum,\cite{spatial}
   role of hadronic modes to bulk plasma properties,\cite{EKR}
 and  $q$-$\bar{q}$ and $q$-$q$ potentials above $T_c$.\cite{NaSa04}
  Further theoretical studies
    on the possible mechanisms\cite{models,DPS01}
 to support  hadronic states above $T_c$ are also called for. 
   
\section*{Acknowledgments}

This work was supported by Grants-in-Aid of the Japanese 
 Ministry of Education, Culture, Sports, Science and Technology, No. 15540254.

\end{document}